\documentclass[12pt]{article}
\usepackage{graphicx}

\newcommand\pubnumber{}
\newcommand\pubdate{\today}

\def\institute{Joint Laboratory of Optics of Palacký University Olomouc and Institute of Physics of Czech Academy of Sciences, Czech Republic}
\def\institutee{Department of Mathematical Analysis and Applications of Mathematics, of Palacký University Olomouc, Czech Republic}
\def\authemail{\footnote{Contact: petr.baron@upol.cz}}
\def\authemaill{\footnote{Contact: jiri.kvita@upol.cz}}
\def\authemailll{\footnote{Contact: radek.privara@upol.cz}}
\def\authemaillll{\footnote{Contact: jan.tomecek@upol.cz}}
\def\authemailllll{\footnote{Contact: rostislav.vodak@upol.cz}}

\def\Title#1{\begin{center} {\Large #1 } \end{center}}
\def\Author#1{\begin{center}{ \sc #1} \end{center}}
\def\Address#1{\begin{center}{ \it #1} \end{center}}

\newcommand\pubblock{\rightline{\begin{tabular}{l} \pubnumber\\
         \pubdate  \end{tabular}}}
\newenvironment{Abstract}{\begin{quotation}  }{\end{quotation}}
\newenvironment{Presented}{\begin{quotation} \begin{center} 
             PRESENTED AT\end{center}\bigskip 
      \begin{center}\begin{large}}{\end{large}\end{center} \end{quotation}}

\def\beq{\begin{equation}}
\def\eeq#1{\label{#1}\end{equation}}
\def\eeqn{\end{equation}}

\def\beqa{\begin{eqnarray}}
\def\eeqa#1{\label{#1}\end{eqnarray}}
\def\eeqan{\end{eqnarray}}

\let\bar=\overbar

\def\Dslash{\not{\hbox{\kern-4pt $D$}}}
\def\dslash{\not{\hbox{\kern-2pt $\del$}}}

\def\msb{{\bar{\ssstyle M \kern -1pt S}}}

\usepackage[font=small,labelfont=bf]{caption} 
\usepackage{url}            

\newcommand{\mindrt}{$\Delta R(J,t)$}
\newcommand{\mindrw}{$\Delta R(J,W)$}
\newcommand{\ttbar}{${t\bar{t}}$}

\newcommand{\pt}{${p_\mathrm{T}}$}
\newcommand{\ptJ}{${p_\mathrm{T}^J}$}

\newcommand{\Delphes}{\textsc{Delphes}}
\newcommand{\Pythia}{\textsc{Pythia}}

\newcommand{\MadGraph}{\textsc{MadGraph5}}

\begin{document}
\begin{titlepage}
\pubblock

\vfill
\Title{Application of Machine Learning Based Top Quark and W Jet Tagging to Hadronic Four-Top Final States Induced by SM as well as BSM Processes}
\vfill
\Author{ Petr Baroň\authemail, Jiří Kvita\authemaill, Radek Přívara\authemailll}
\Address{\institute}
\Author{ Jan~Tomeček\authemaillll, Rostislav Vodák\authemailllll}
\Address{\institutee}
\vfill
\begin{Abstract}
We apply gradient boosting machine learning techniques to the problem of hadronic jet substructure recognition using classical subjettiness variables available within a common parameterized detector simulation package DELPHES. Per-jet tagging classification is being explored. Jets produced in simulated proton-proton collisions are identified as consistent with the hypothesis of coming from the decay of a top quark or a W boson and are used to reconstruct the mass of a hypothetical scalar resonance decaying to a pair of top quarks in events where in total four top quarks are produced. Results are compared to the case of a simple cut-based tagging technique for the stacked histograms of a mixture of a Standard Model as well as the 
new physics process.
\end{Abstract}
\vfill
\begin{Presented}
$16^\mathrm{th}$ International Workshop on Top Quark Physics\\
(Top2023), 24--29 September, 2023
\end{Presented}
\vfill
\end{titlepage}
\def\thefootnote{\fnsymbol{footnote}}
\setcounter{footnote}{0}

\section{Introduction}
Machine learning (ML) techniques are getting growing application in many research areas, one of them being the events classification in high energy physics (HEP). 
The structure of this paper is as follows.
We first study the application of selected gradient boosting ML techniques to the recognition of a substructure of hadronic final states (jets) and their tagging based on their possible origin in current HEP experiments using simulated events and a parameterized detector simulation. We present the samples, their jet composition, and results of tagging in both per-jet and per-event cases. We then check independently the tagging efficiencies and apply the trained taggers to dedicated samples with a clear signature of a jet mass peak. We also compare to standard cut-based methods and compare the physics performance as well as correlations between the classical and ML based tagging methods, with implications to searches for hadronically decaying four-top final states in proton-proton collisions.

\section{Objects}
Using the \MadGraph{} version {\tt 2.6.4} simulation toolkit~\cite{Alwall:2014hca}, proton-proton collision events at $\sqrt{s} = 14$ TeV were generated for the SM process $pp \rightarrow$ \ttbar{} in the all-hadronic \ttbar{} decay channel at next-to-leading order (NLO) in QCD in production, using the MLM matching~\cite{Hoeche:2005vzu,Mangano:2002ea}, \emph{i.e.} with additional processes with extra light-flavoured jets produced in the matrix element, matched and resolved for the phase-space overlap of jets generated by the parton shower using \MadGraph{} defaults settings. The parton shower and hadronization were simulated using \Pythia{}8~\cite{Sjostrand:2014zea}. As a train BSM model, the resonant $s$-channel \ttbar{} production via an additional narrow-width (sub-GeV) vector boson $Z'$ as $pp \rightarrow Z' \rightarrow$ \ttbar{} using the model \cite{FeynModelZprime,Christensen:2008py,Wells:2008xg}
were generated, to provide a sample of top quarks with large transverse momenta, enhancing the boosted regime. As a representative model of a BSM process for testing, the production of a scalar resonance decaying to a pair of top quarks $y_0 \rightarrow$ \ttbar{} was adopted~\cite{Christensen:2008py} at the leading-order (LO) in the \ttbar{} production with the gluon-gluon fusion loop 
, with inclusive \ttbar{} decays, selecting the all-hadronic channel later in the analysis. 
\subsection{Parameterized detector simulation}
Using the \Delphes{} (version {\tt 3.4.1}) detector simulation \cite{deFavereau:2013fsa} with the ATLAS card, jets with distance parameters of $R=1.0$ (dubbed as large-$R$ jets) were reconstructed using the anti-$k_t$ algorithm using the FastJet package~\cite{Cacciari:2011ma} at both particle and detector levels. The trimming jet algorithm \cite{Krohn:2009th} as part of the \Delphes{} package was used to obtain jets with removed soft components, using the parameter of $R_\mathrm{trim} = 0.2$ and modified \pt{} fraction parameter $f^{{p_\mathrm{T}}}_\mathrm{trim} = 0.03$ (originally 0.05). The trimming algorithm was chosen over the standard non-groomed jets, soft-dropped~\cite{Larkoski:2014wba} and pruned jets~\cite{Ellis:2009me}, with parameters varied, in terms of the narrowness of the mass peaks.
\subsection{Objects of interest}
The interest is the identification of large-$R$ hadronic jets coming from the hadronic decays of top quarks and $W$ bosons. In the na\"{i}ve picture of the hadronic decays of $W \rightarrow q\bar{q}'$ and $t \rightarrow W b \rightarrow b q\bar{q}'$, these three and two prong decays, respectively. Different jet substructure is thus expected for such $t$ and $W$ jets.

\subsection{Cut-based tagging}
As input variables to both cut-based as well as ML-based tagging we utilize simple yet powerful ``classical'' variable called $n$-subjettiness~\cite{Thaler_2011}, $\tau_N$, which is related to the consistency of a jet with the hypothesis of containing $N$ subjets.
These variables are combined into ratios $\tau_{32}$ and $\tau_{21}$, defined as $\tau_{ij} \equiv \frac{\tau_i}{\tau_j}$. 
In order to identify jets coming from the hadronic decays of the $W$ boson or a top quark by a simple cut-based algorithm, large-$R$ jets were tagged as
\begin{itemize}
\itemsep-0.25em 
\item $W$-jets if $ 0.10 < \tau_{21} < 0.60 \, \land \,   0.50 < \tau_{32} < 0.85 \, \land \,  m_J \in [70, 110] \,\mathrm{GeV}$;
\item top-jets if $ 0.30 < \tau_{21} < 0.70 \, \land \,   0.30 < \tau_{32} < 0.80 \, \land \,  m_J \in [140, 215] \,\mathrm{GeV}$.
\end{itemize}

Shapes of the variables used as input to the ML classifier are shown in Figure~\ref{fig:mass_taus_samples} for the individual samples.
One can observe the enhancement in the $Z'$ samples at the place of the expected top quark mass peak, the larger the higher the mass of the $Z'$ particle, while the lower mass $Z'$ sample provides enhanced region at the $W$ boson mass. The various \ttbar{} samples exhibit a large continuum of masses, with non-resonant bulk contribution below 60~GeV of different sizes due to different jet \pt{} kinematics cut for the samples.
\begin{center}\vspace{-0.3cm}
    \includegraphics[width=0.4\linewidth]{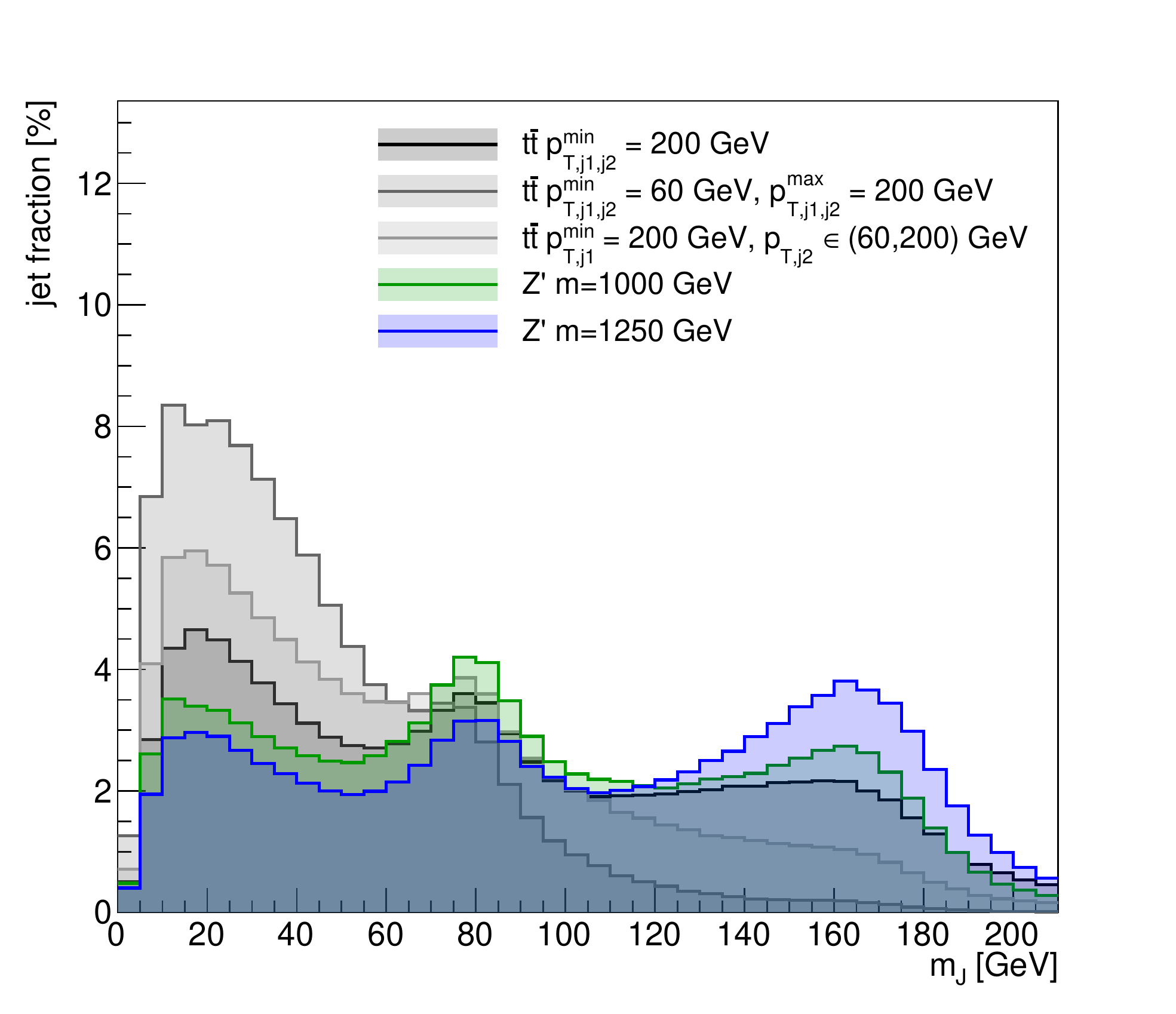}
    \captionof{figure}{Shapes of the large-$R$ jet mass variables in the five samples for training and testing.}
    \label{fig:mass_taus_samples}
\end{center}
\section{ML-based top and $W$ tagging}
\subsection{General data structure}
Three samples corresponding to the SM \ttbar{} production were generated, with different cuts at the generator level on the transverse momentum of the jets, in order to cover regions with various fractions of $t$, $W$ as well as non-resonant (light) jets. These have been used as both training and testing data sets. The two $Z'$ samples with the $Z'$ masses of $1000$ and $1250\,$GeV provide a \ttbar{} sample with enhanced boosted top quarks, thus leading to events with enhanced fractions of $t$ and $W$ jets.
Variables defined and used for each jet in the classification are as follows
\begin{itemize}
    \itemsep-0.25em 
    \item \mindrw{}, the minimal distance of the jet to the nearest~$W$~\footnote{The angular distance between two objects is defines as $\Delta R \equiv \sqrt{ (\Delta\phi)^2 + (\Delta\eta)^2}$ where the pseudorapidity $\eta \equiv -\ln\tan\frac{\theta}{2} $ is related to the standard azimuthal angle $\theta$ of the spherical coordinates, where the beam axis coincides with the $z$ axis, and $\phi$ is the polar angle in the $xy$ plane.}; 
    \item \mindrt{}, the minimal distance of the jet to the nearest top parton; 
    \item Jet tr. momentum \ptJ ~and jet four-vector invariant mass~$m_J$.
    \item $\eta$  and $\phi$ of the jet.
    \item Jet substructure variables $\tau_{32}$ and $\tau_{21}$.
\end{itemize}
\noindent
The true type jets labels are based on the following criteria
\begin{enumerate}
\itemsep-0.25em 
\item truth $t$-jets:  \mindrt{}$<0.25 \land 150\,\mathrm{GeV}\leq m_J \leq 210\,\mathrm{GeV}$;
\item truth $W$-jets; \mindrw{}$<0.25 \land 60\,\mathrm{GeV}\leq m_J \leq 110\,\mathrm{GeV}$;
\item truth light jets: otherwise.
\end{enumerate}
For the predictions we used the machine learning (ML) model based on the Gradient Boosting technique, which is a popular 
 and widely used algorithm for supervised learning, see e.g.~\cite{MullerGuido}. This classifier is one of the two most 
 used types of \emph{ensemble methods}, which are methods combining  multiple simple predictors (esp. decision trees) in 
 order to create a more powerful model. The method does not work with weights but it tries to fit the predictor to the 
 \emph{residual errors} made by the previous predictor. The new prediction is made by simply adding up the predictions of 
 all the predictors. We decided to test per-jet predicting of t-, W- or l-jets.

 \subsection{Results}

 Figure~\ref{fig:mass_taus_samples2} displays the jet mass distribution, where the red filled areas represent top-tagged jets using both the cut-based (dark red) and ML approach (light red). 
 Although the cut-based method exhibits higher tagging efficiency, it is evident that it suffers from a significant mistagging of light jets as tops. 
 In contrast, the ML method appears to tag top jets with higher purity, and the boundary of the red filled area aligns with the background of light jets. 
 To further support this argument, the real tagging efficiencies as a function of jet $p_{T}$ are presented in Figure~\ref{fig:mass_taus_sample4}, 
 where the ML approach demonstrates lower efficiency in all bins (solid red) compared to the cut-based approach (dashed red). 
 Additionally, the ML method exhibits lower fake efficiency in the QCD background (blue lines).
 Figure~\ref{fig:mass_taus_sample4} showcases the invariant mass of leading and subleading top-tagged jets for the Standard Model $t\bar{t}t\bar{t}$ process, 
 alongside the stacked invariant mass of leading and subleading jets from the $y_{0}t\bar{t}\rightarrow t\bar{t}t\bar{t}$ process (scaled by a factor of 0.075). 
 The light filled areas and dashed lines represent the top-tagged jets that are matched to the parton top within $\Delta R < 0.15$. 
 The efficiencies, as indicated in the legend of Figure~\ref{fig:mass_taus_sample4}, are provided for each set of histograms as 
%
%
\begin{equation}
  \epsilon = \frac{\mathrm{top~tagged~jets~}\&\&\mathrm{~matched~to~parton~level}}{\mathrm{top~tagged~jets}}.
\end{equation}
\begin{center}\vspace{-0.5cm}
    \includegraphics[width=0.48\linewidth]{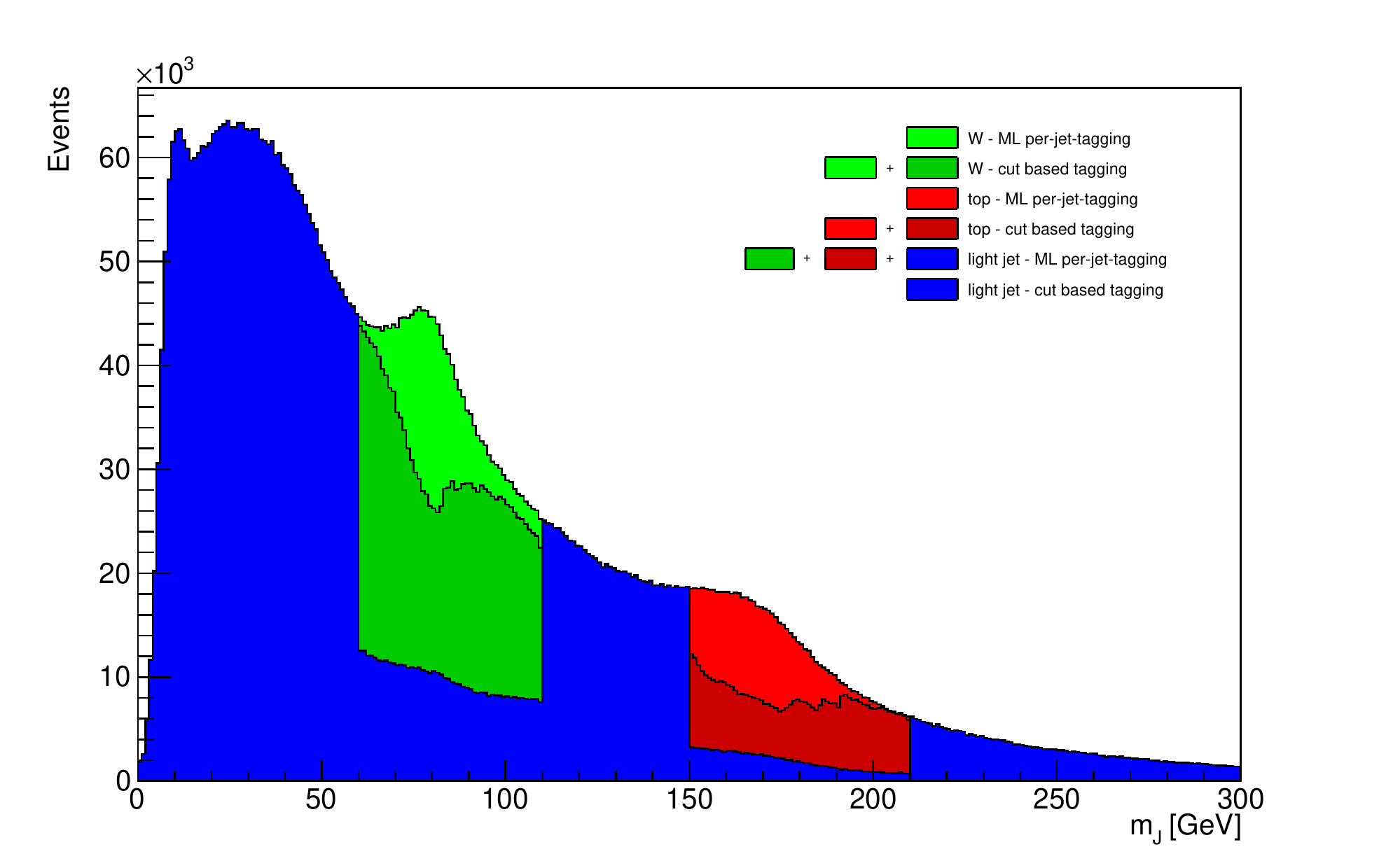}
    \includegraphics[width=0.48\linewidth]{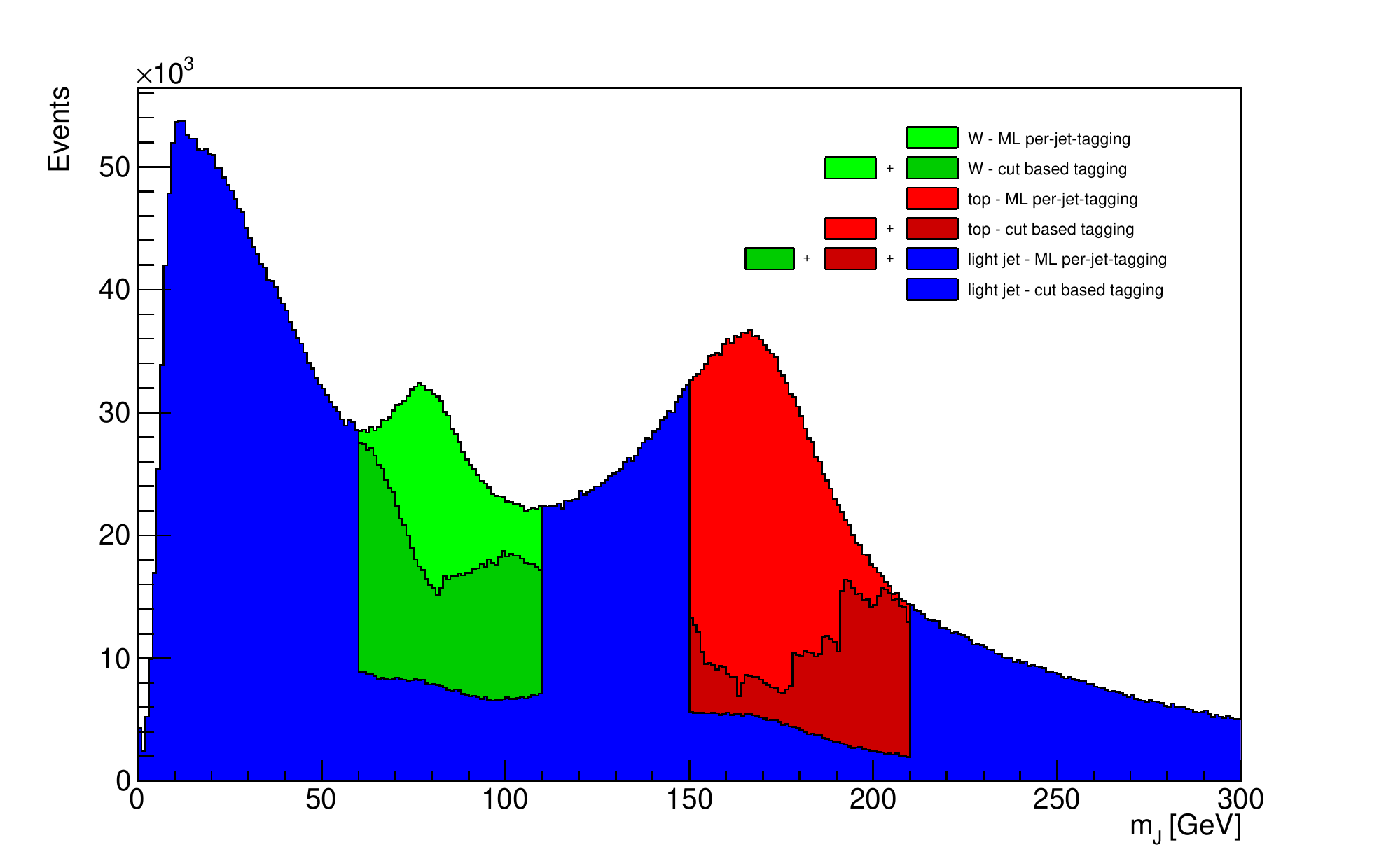}
    \captionof{figure}{Jets mass of Standard Model $t\bar{t}t\bar{t}$ (left) and $y_{0}t\bar{t}\rightarrow t\bar{t}t\bar{t}$ (right) with cut-based and maching learning tagging of $W$ boson (green) and top (red).}
    \label{fig:mass_taus_samples2}
\end{center}

\begin{center}
  \includegraphics[width=0.48\linewidth]{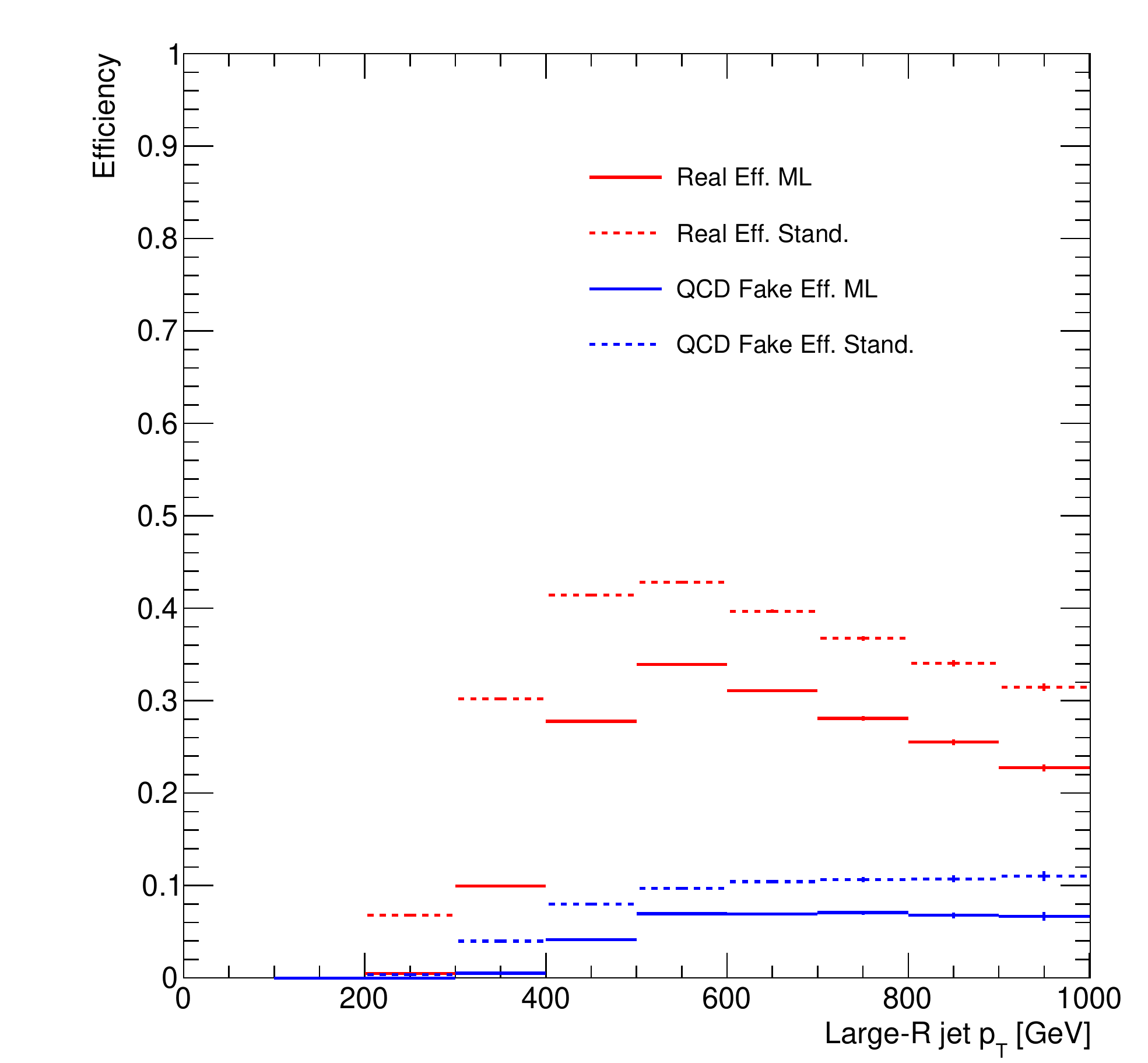}
  \includegraphics[width=0.48\linewidth]{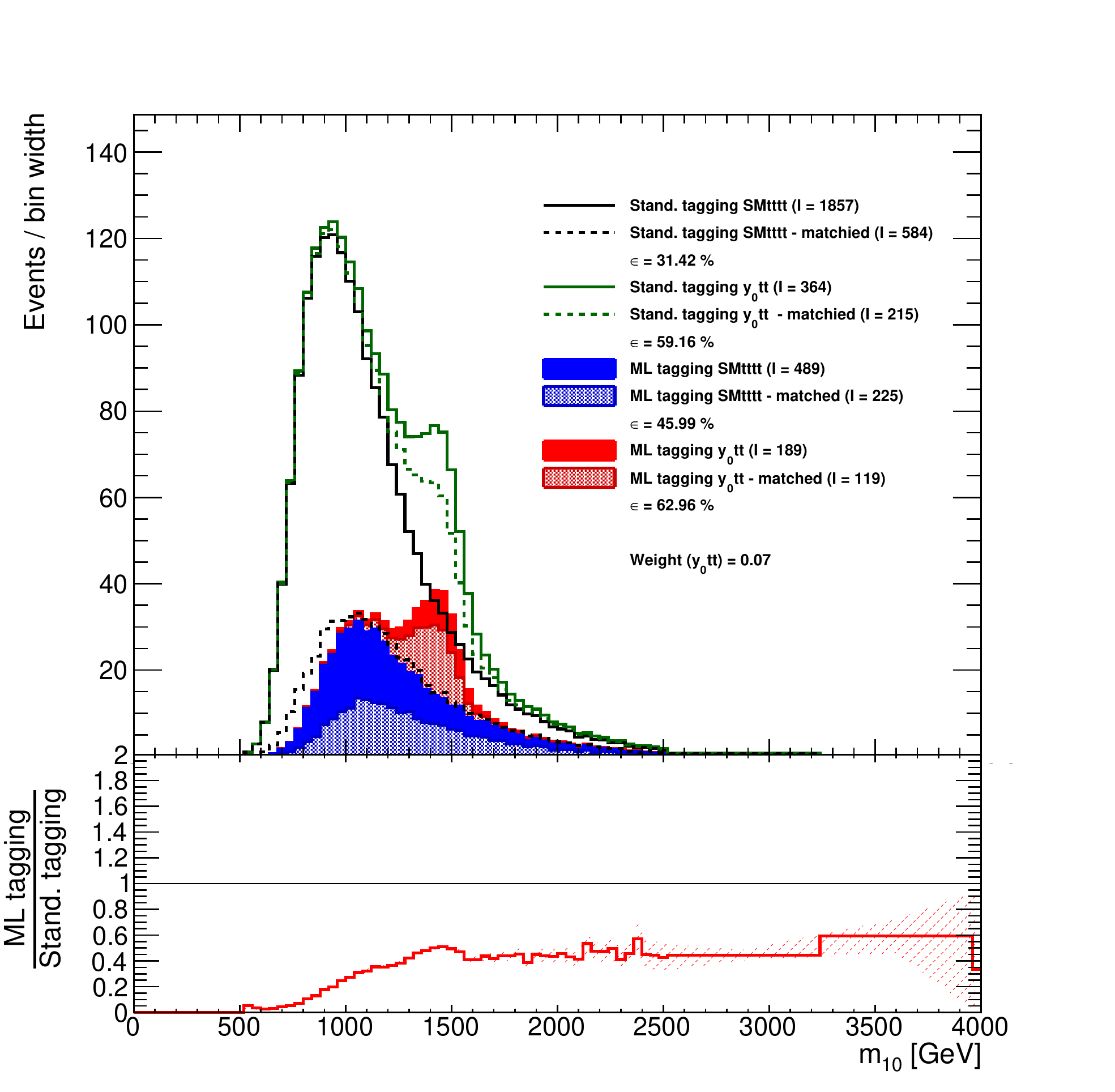}
  \captionof{figure}{{\bf Left: }Real efficiencies of the top-tagging using cat-based (dashed red line) and ML approach (solid red line) and QCD background fake efficiencies using cut-based (dashed blue lines) and ML (solid blue line). {\bf Right: }Invariant mass of leading and subleading jets of Standard Model $t\bar{t}t\bar{t}$ and $y_{0}t\bar{t}\rightarrow t\bar{t}t\bar{t}$ ($m_{y_{0}} = 1.5~\mathrm{TeV}$) tagged as top quark. The light red / blue filled areas and dashed lines represent matched jets to top quark within $\Delta R < 0.15$.}
  \label{fig:mass_taus_sample4}
\end{center}

\section*{Conclusion}
The ML approach demonstrates a higher tagging efficiency compared to top-tagged jets matched at the parton level, 
with an improvement of 14.6\% for the Standard Model $t\bar{t}t\bar{t}$ process and 3.8\% for the $y_{0}t\bar{t}\rightarrow t\bar{t}t\bar{t}$ process 
(with $m_{y_{0}} = 1.5~\mathrm{TeV}$) when compared to the cut-based selection.

\section*{Acknowledgement}
The authors gratefully acknowledge the support from the project IGA\_PrF\_2023\_005 of Palacky University. 
The authors would like to thank the grants of MSMT, Czech Republic, GACR 19-21484S, and GACR 23-07110S for the support.
Initial work and classifiers were supported by the project GACR 19-21484S while the final results and travel by GACR 23-07110S.



\end{document}